\documentstyle[preprint,pre,aps]{revtex}
\tighten
\begin{document}
\draft

\title{Directed percolation conjecture
for cellular automata revisited }

\author{G\'eza \'Odor and Attila Szolnoki}

\address
{\it Research Institute for Materials Science, P.O.Box 49,
H-1525 Budapest, Hungary }

\date{\today}
\maketitle

\begin{abstract}
The directed percolation (DP) hypothesis for stochastic,
range-4 cellular automata with acceptance rule
$y \le\sum_{j=-4}^4 s_{i-j} \le 6$, in cases of $y < 6$ was
investigated in one and two dimensions.
Simulations, mean-field renormalization group and
coherent anomaly calculations show that in one dimension the
phase transitions for $y<4$ are continuous and belong to the
DP class for $y=4,5$ they are discontinuous.
The same rules in two dimensions for $y=1$ show $2+1$
dimensional DP universality; but in cases of $y > 1$ the
transitions become first order.
\end{abstract}


\narrowtext

\section{Introduction}

The phase transition theory of nonequilibrium systems is
an area of considerable interest.
There is not such a clear view about critical universality
classes as there is for equilibrium systems.
The cellular automata (CA) approach for modelling
nonequilibrium systems can be regarded as a general tool like
differential equations. It introduces a space and time
discretization --- thus making the solution treatable --- but
preserving the features of interest. As was shown in \cite{Wolf}
CA models can describe very rich behavior. Even the simplest
one dimensional CA can have various phase transitions, can
generate chaotic patterns, or show complex self-organizing or
reproductive behavior.

The directed percolation (DP) hypothesis \cite{DP} states
that all continuous transitions of a scalar order parameter
to a unique absorbing state in dimension $d$ belong to the DP
universality class, represented by the $d+1$ dimensional directed
percolation \cite{Cardy} or reggeon field theory.
There are some additional conditions -- usually not stated
explicitly -- to be fulfilled as well :
short range interaction both in space and time, translational
invariance, absence of multicritical points, and
the non-vanishing probability for any active state to die locally.
Although there is no proof for this conjecture there exists
a great number of models, which have been found to belong to
this class. Attempts to find counterexamples \cite{Taka,Bid,Alb}
by other simulation results \cite{Jensen} have been unsuccessful.
The motivation of our investigation is to find a possible
exception to the DP conjecture in the cellular automata
phase transition of different dimensions.
The stochastic cellular automata -- investigated by Bidaux,
Boccara and Chat\`e (hereafter referred to as BBC CA)
\cite{Bid} and Jensen \cite{Jensen} using simulations -- in one
dimension showed controversial results concerning the universality
class. Steady state simulation gave non-DP whereas the time-dependent
simulation resulted in the DP class. Even though the latter method is
considered to be more precise, arguments based on universality
suggested that the same rule in two dimensions should produce
a transition of 2+1 dimensional DP class. Simulations proved this to
be a first order transition \cite{Bid}. In order to clarify the
situation we used various selection of different analytical and
simulation methods.

In addition we were interested in the change of universality
by varying the lower acceptence value $y$ of the rule :

\[ s(t+1,j) = \left\{ \begin{array}{ll}

X & \mbox{if} \ \ y \leq \sum_{j-4}^{j+4}s(t,j) \leq 6 \\
0 & \mbox{otherwise \ \ ,}
\end{array} \right. \]
where $X\in\{0,1\}$ is a two valued random variable such that

$$Prob(X=1)=p \ \ .$$

By setting up the mean-field steady state equation for the
probability of ones (c) :
\begin{equation}
{\partial c\over\partial t} = f(p,c,t) = 0
\end{equation}
the leading term in the polinomial function $f$, which determines
the critical behavior as $p\to p_c$ (and $c\to 0$) is $c$
for $y=1$, $c^2$ for $y=2$ and $c^3$ for $y=3$.
In analogy with the equilibrium Ising model --- where the
three-spin Ising model has a different universality from the
two-spin nearest neighbor Ising model --- we expected deviations
from the DP behavior.
To check this assumption we extended our investigation to
the $y=1,2,3$ cases in one and two dimensions.

\section{Steady state simulations}

The steady state Monte Carlo simulations (STDMC) were carried
out on $L = 40000$ and $L = 256 \times 256$ sized lattices with
periodic boundary conditions. In each case measurements at
$30$--$50$ different $p$ values were taken. Following
$10^5$ -- $5\cdot 10^5$ relaxation lattice updates at each $p$
the concentration averages and r.m.s. values were determined.
The exponent $\beta$ and $p_c$ were fitted together -- using
the weighted least squares error method -- to the critical
scaling formula
\begin{equation}
c(p) \sim \vert p - p_c\vert^{\beta} \ . \label{ordp}
\end{equation}

Since the concentration in the immediate neighborhood of the
critical point has a relatively large statistical error
-- owing to the increased relaxation time -- data points
within the $1 \%$ neighborhood of $p_c$ were discarded.
In order to compensate this truncation of data
we performed the fitting in several different neighborhoods
of the $p_c$. Since we don't know how the (\ref{ordp}) scaling
breaks down away from $p_c$, we have checked this tendency
of the fitting by varying $p_{cut}$ of the used data points :
$(p_c\times 1.01) <  \{p\}  < p_{cut}$.
This kind of analysis enabled us to see how the critical
exponent converged as $p_{cut} \to p_c$ and we were
able to extrapolate for the true $\beta$ exponent in spite
of the growing statistical errors in the vicinity of $p_c$.

\subsection{Results in one dimension}

The generally accepted 1+1 dimensional DP critical exponets
are summarized in Table \ref{DP}.
Figure \ref{fig1} shows that the estimated critical exponent of
the order parameter converges to the value $\beta\approx 0.28$,
which is about the DP value determined by Dickman
and Jensen by series expansion \cite{Dickj}.
The critical point and $\beta$ estimates ( given in Table
\ref{pcbeta}) were determined by linear weighted least-square
fitting to the functions of $p_{cut}$.
The results of the BBC \cite{Bid} ($y=3$) model are the
most uncertain, but still seem to extrapolate to the DP
class value if $p_{cut}\to p_c$.
Estimates based on fitting with less than $10$ data points
should not be taken very seriously and are shown for
information only.

As a tendency the transitions become more and more
steep (and $p_c\to 1$) as we increase the lower limit of the
CA rule ($y$). For $y=4$ and $5$ non-empty steady state exists in
the deterministic, $p=1$ limit only. One can easily check
that only a few oscillating structures can survive. There is a
first order transition to the empty state if $p < 1$, which is
quite rare among one dimensional CA. For $y>5$ every cell dies out.

To verify our results further simulations and analytical
calculations were carried out; these are discussed in the subsequent
sections.

\subsection{Results in two dimensions}

In two dimensions for $y=1$ the transition seems to
be continuous and the estimates for the critical exponent
converged to the $2+1$ dimensional DP class value :
$\beta\approx 0.58$ \cite{Brower} (Figure \ref{fig1}/d).
For $y=2$ and $3$ the transitions appear to be discontinuous,
see Figures \ref{fig2}. If formula (2) is fitted to the
concentration data -- as if continuous
transitions were assumed -- results in $p_c$-s with values lower
than $p$ with non-zero densities.
Since it is not easy to prove first order transition by simulation,
we applied the generalized mean-field method,
which has already been found to be very useful for deciding
the order of the transition at CA \cite{OdBocSza}. The results
of these calculations are also shown in Figure \ref{fig2}
and they are discussed in section \ref{sec:GMF}.

\subsection{Finite-size scaling analysis}

Finite-size scaling is shown to be applicable
to continuous transition to an absorbing state
of nonequilibrium systems \cite{Auk}. At the critical
point the steady state density ($c$) and the fluctuation
$\chi = L^d (<c^2> - <c>^2)$ scale with the system size as :
\begin{equation}
c(L) \propto L^{-\beta/\nu_{\perp}},  \label{cs}
\end{equation}
\begin{equation}
\chi (L) \propto L^{-\gamma/\nu_{\perp}}, \label{fs}
\end{equation}
where $\nu_{\perp}$ is the correlation length exponent
in the space direction :
\begin{equation}
\xi(p)\propto \vert p - p_c \vert^{-\nu_{\perp}}.
\end{equation}
Simulations were done in one dimension for lattice
sizes : $L = 16,32,64,...,8192$ with
the corresponding timesteps : $t = 200,400,800, ... , 102400$.
Averaging was performed on the 'surviving' samples out of
$N_s = 262144, 131072, 65536, ..., 512$ independent runs.
The $p_c$ values are taken from the time-dependent MC calculations.
Figure \ref{fig3} shows that the $c$ and $\chi$ results fall nicely
on a straight line on a log-log plot, but for $y=1$ the fitting
resulted in:
$$\beta/\nu_{\perp} = 0.260(4), \ \ \ \gamma/\nu_{\perp} = 0.43(1),$$
for $y=2$ :
$$\beta/\nu_{\perp} = 0.252(1), \ \ \gamma/\nu_{\perp} = 0.548(7),$$
and for $y=3$ :
$$\beta/\nu_{\perp} = 0.258(1), \ \ \gamma/\nu_{\perp} = 0.496(4),$$
which differ slightly from the DP class values \cite{Jen}:
$$\beta/\nu_{\perp} = 0.2522(6), \ \ \gamma/\nu_{\perp} = 0.496(2).$$

We have shown that the deviations come from systems with small sizes,
similarly as for \cite{Jen}. By omitting points with sizes less than
$L = 16,32,64$ the results converge to the DP values
(see Tables \ref{fss1_6}, \ref{fss2_6}, \ref{fss3_6})
and we can see the emergence of a DP like behavior in the large $L$
limit.

\section{Time-dependent simulations}

Time-dependent Monte Carlo simulations (TDMC) have been shown to be an
efficient
method for locating critical points and estimating exponents for
models with continuous transition to an absorbing state
\cite{Jen,Gras89,Dick90}. The initial state of this
simulation is the smallest a single live 'seed', that can survive
at the center of the lattice. The evolution of the population is
followed up to several thousands of lattice update steps and
the following quantities are measured at each time step :
\begin{itemize}
\item survival probability $s(t)$
\item concentration $c(t)$
\item average mean square distance of spreading from the
      center $R^2(t)$
\end{itemize}
The evolution runs are averaged over $N_s$ independent runs
for each different value of $p$ in the vicinity of $p_c$ ( but
for $R^2(t)$ only over the surviving runs ).
At the critical point we expect these quantities to behave
in accordance with the power law as $t\to\infty$, i.e.
\begin{equation}
s(t)\propto t^{-\delta} \ ,
\end{equation}
\begin{equation}
c(t)\propto t^{\eta} \ ,
\end{equation}
\begin{equation}
R^2(t)\propto t^z \ .
\end{equation}
We carried out simulations for the one-dimensional systems
up to $t = 5000$ time steps with $N_s \simeq 10^6$ independent
samples. The seed was a single $1$ for $y=1$, a neighboring
pair of $1$-s for $y=2$, and a triplet for $y=3$.
We did not follow the evolution of $R^2(t)$.
To estimate the critical exponents we determined the local slopes :
\begin{equation}
-\delta(t) = {\ln \left[ s(t) / s(t/m) \right] \over \ln(m)}
\end{equation}
\begin{equation}
\eta(t) = {\ln \left[ c(t) / c(t/m) \right] \over \ln(m)}
\end{equation}
using $m=8$. Figures \ref{fig4}, \ref{fig5}, \ref{fig6} show these
quantities as functions of $t^{-1}$.
In the case of power-law behavior we should see
a straight line as $t^{-1} \to 0$, when $p = p_c$. The off-critical
curves should possess curvature. Curves corresponding to $p > p_c$
should veer upward, curves with $p < p_c$ should veer down.
The critical exponent should be read off as the interception
of the critical (straight) curve with the ordinate axis.

It is emphasized that it is very difficult to get precise estimates
on the critical data by visual inspection of the curves . While for
$y=3$ the middle curve -- which seems to be the nearest to the
critical one -- one can read off DP exponents, for $y=1,2$ one
could conclude that we have non-DP values.
The logarithmic $c(t)$ and $s(t)$ data were fitted by parabolas :
\begin{equation}
\ln(c(t)) = const. + \eta_1 \ln(t) + \eta_2 \ln^2(t)
\end{equation}
\begin{equation}
\ln(s(t)) = const. + \delta_1 \ln(t) + \delta_2 \ln^2(t)
\end{equation}
at different $p$ values. Estimations for $p_c$ are based
on the conditions $\eta_2 = 0$ and $\delta_2 = 0$;
the results are shown in Table \ref{pcbeta}.
It should be noted, however that even though the difference
is small ($\Delta p < 10^{-4}$) the conditions $\eta_2 = 0$
and $\delta_2 = 0$ do not give exactly the same $p_c$.
Also the estimates for the exponents are slightly off the DP
class values.
The errors must have statistical nature with finite size
errors being excluded since the fronts of the populations ---
over the entire run --- cannot exceed the boundary of the lattice.

To get more precise exponents we used the general form
-- including corrections to scaling -- of the local slopes
\cite{Gras89}
\begin{equation}
\delta(t) = \delta + a t^{-1} + b \delta^, t^{-\delta^,} + ...
\end{equation}
where $\delta^,$ is a correction to the scaling exponent.
As indicated in Table \ref{pcbeta} the exponents agree well
with the DP class data.

\section{Mean-field renormalization group analysis}

Renormalization group methods for nonequilibrium
systems have been successful when combined with mean-field
approximation. At the critical point the scaling behavior
of quantities calculated on small clusters of different
sizes is exploited. One main branch of such calculations
is the Coherent Anomaly Method (CAM) its application
to these CA models will be discussed in section \ref{sec:CAM}.
Another important approach is the so-called
mean-field renormalization group (MFRG) method, originally
introduced by Indekeu et al. \cite{Indekeu} for
equilibrium systems. Later it was shown that the method
may also be applied to self-avoiding walk and percolation
problems \cite{Debell}. The application to nonequilibrium systems
was demonstrated by \cite{Marques}. Although this method is
not considered to be the most accurate, it is relatively
simple and effective. Developments to enhance its
performance are under way \cite{Croes}. Here, we utilize the
approach for one-dimensional CA.

First we set up mean-field equations for different sized
block probabilities ($P_i$) in the steady state. Here the
index $i$ denotes an $n$ block.
However we do not use the Bayesian extension trick as
in the case of the generalized mean-field calculations,
but take into account the external
field with the average concentration ($c$):
\begin{equation}
0 = {\partial P_i(p)\over\partial t } = f( \{P_j(p) \},c)
\end{equation}
Because in the vicinity of the second-order phase transition to
the absorbing state $c\to 0$, we expand and keep linear
terms only in the equations. We express the probability of
a site being occupied (${\it O}^n(p,c)$) by calculating it
in terms of different ($n$) cluster sizes.
For the $y=1$ model in the $n=1$ case the equations are :
\begin{equation}
-8\,c\,p\,{\it P_0} + \left( 1 - 8\,c \right) \,\left( 1 - p \right) \,
      {\it P_1} = 0,
\end{equation}
\begin{equation}
   {\it P_0} + {\it P_1} = 1.
\end{equation}
In view of this, the probability of a site being in state 1 is :
\begin{equation}
   {\it O}^1(p,c) = {{8\,c\,p}\over {1 - 8\,c - p + 16\,c\,p}} .
\end{equation}
Expanding in factors of $c$ we get :
\begin{equation}
{\it O}^1(p,c) = {{8\,c\,p}\over {1 - p} } + {\cal O}(c^2).
\end{equation}
For clusters of pairs ($n=2$) the equations look like :
\begin{eqnarray*}
     0 & = & -2\,c^,\,p\,{\it P_{00}} - 12\,c^,\,\left( 1 - p \right)
\,p\,{\it P_{00}}-
     6\,c^,\,{p^2}\,{\it P_{00}} \\ & + &
     2\,\left( 1 - 8\,c^, \right) \,{{\left( 1 - p \right) }^2}\,{\it P_{01}}
\\ & + &
     \left( 1 - 8\,c^, \right) \,{{\left( 1 - p \right) }^2}\,{\it P_{11}},
\end{eqnarray*}
\begin{eqnarray*}
     0 & = & 2\,c^,\,p\,{\it P_{00}} + 12\,c^,\,\left( 1 - p \right)
 \,p\,{\it P_{00}} \\ & - &
     2\,\left( 1 - 8\,c^, \right) \,{{\left( 1 - p \right) }^2}\,{\it P_{01}} -
     2\,\left( 1 - 8\,c^, \right) \,{p^2}\,{\it P_{01}}  \\ & + &
     2\,\left( 1 - 8\,c^, \right) \,\left( 1 - p \right) \,p\,{\it P_{11}},
\end{eqnarray*}
\begin{eqnarray*}
     0 & = &6\,c^,\,{p^2}\,{\it P_{00}} +
     2\,\left( 1 - 8\,c^, \right) \,{p^2}\,{\it P_{01}} \\ & - &
     \left( 1 - 8\,c^, \right) \,{{\left( 1 - p \right) }^2}\,{\it P_{11}} \\ &
- &
     2\,\left( 1 - 8\,c^, \right) \,\left( 1 - p \right) \,p\,{\it P_{11}},
\end{eqnarray*}
\begin{equation}
   {\it P_{00}} + 2\,{\it P_{01}} + {\it P_{11}} = 1,
\end{equation}
Now, the probability of 1 after linearization is :
\begin{equation}
{\it O}^2(p,c^,) = {{c^,\,p\,\left( 7 + {p^2} \right) }\over
{{{\left( -1 + p \right) }^2}}} + {\cal O}(c^2).
\end{equation}
At the critical point the scaling requires that the ${\it O}^n(p,c)$
probability must scale like the external field :
\begin{equation}
{{\it O}^{n^,}(p,c^,) \over {\it O}^n(p,c)} = {c^, \over c}.
\end{equation}
This condition enables one to combine the expressions for ${\it O}^n$ and
${\it O}^{n^,}$ and to get an estimate for the $p_c$. Comparing the $n=1$
and $n=2$ results we obtained $p_c(1,2) = 0.123106$.

The exponent $\nu_{\perp}$, which measures the critical behavior
of the correlation length, can be obtained from the scaling of $(p-p_c)$
resulting from change of the length scale $l = {n^, / n}$ :
\begin{equation}
{\partial {\it O}^{n^,}(p,c^,) \over \partial {\it O}^n(p,c)}\bigg|_{p=p_c}
 = l^{1/\nu_{\perp}}
\end{equation}
For $n=1$ and $n=2$ clusters it is : $\nu_{\perp}(1,2) =  5.802$.
This is very far from the DP class $\nu_{\perp}\sim 1.1$ value, but
if calculations are done on larger clusters the results converge there.
We were able to go up to $n=5$ clusters with the help of symbolic
MATHEMATICA programming. Table \ref{mfrg} shows that there is a slow
convergence of the $p_c$ and $\nu_{\perp}$ values.
With increasing the cluster sizes we expect more accurate
results.

Unfortunately we were not able to apply the method for $y=2,3$
models because the mean-field approximations predict first order
transitions for small cluster sizes and thereafter no scaling
is expected.

\section{Generalized mean-field results}
\label{sec:GMF}

The generalized mean-field approximation (GMF) introduced by
\cite{Gut,Dick88} has already been applied to describe phase
transitions of different stochastic CA models \cite{SzaOd,OdBocSza}
with success. The phase structure and the tricritical point
of the one-dimensional BBC model ($y=3$) with site-exchange
mixing has been explored with the help of GMF \cite{OdBocSza}.

The traditional mean-field approximation has been extended
to handle clusters of sizes $n>1$ with the help of the
Bayesian extension. Correlations of range $> n$ at a given
level of approximation ($n$) are neglected and the convergence
of the results is analysed by increasing $n$; the details of
the technique are given in  \cite{SzaOd}. The extension of
the method to the two-dimensional CA model is straightforward.
In this case the nearest- and next-nearest neighbor of the
lattice point can contribute to the sum defined in (1).
Taking into account the probability of a given distribution we
used maximal overlap of the clusters. At three-point level the
form of the cluster breaks the $x-y$ symmetry and prevents
the configuration from being covered symmetrically. Because of
this the result of the equations depends on the way of covering the
configuration.
To avoid this uncertainty we restricted ourselves to symmetrical
clusters ($n=$ 1, 2, 4).

\subsection{Results in one dimension}

In the case of the $y=1$ model the convergence of the
approximations is monotonic. Even the traditional ($n=1$)
mean-field approximation predicts continuous phase transition.
By applying quadratic fitting to the $p_c(n)$ results we got
an extrapolation value of $n\to\infty$ value, which agrees
with the simulation result (see Table \ref{GMF1}).
For $y=2,3$ the low level approximations gave a discontinuous
phase transition, but the gap sizes decrease with increasing $n$
and finally disappear at $n=6$ for $y=2$, and at
$n=3$ for $y=3$. It is interesting that the convergence is
non-monotonic for the $y=3$ case in contrast to
what we expect for equilibrium systems.

\subsection{Results in two dimensions}

As one expects in two dimensions the GMF approximation gives
more precise results than in one dimension. For the $n=4$ level
approximation the $p_c$ data differ from the simulations by
a few per cent only. For the $y=1$ model all approximations
predict a second order phase transition, whereas for $y=2,3$
cases the transitions are discontinuous (see Table \ref{GMF2})
and the gap sizes increase with $n$. We can expect that
in the $n\to\infty$ exact limit the transition remains first
order in agreement with the STDMC simulations.

\section{Coherent anomaly extrapolation}
\label{sec:CAM}

The GMF approximations give slow convergence at the critical
point since the correlation length there goes to infinity.
Therefore the accuracy of method breaks down in the
immediate vicinity of $p_c$. To obtain estimates for the
critical exponents one needs to apply an extrapolation technique.
This technique usually is based on finite size scaling theory.
The mean-field renormalization technique is one such approach.

The coherent anomaly method (CAM) introduced by Suzuki
\cite{Suz} is based on the scaling relation that the solution
for singular quantities at a given ($n$) level of
approximation ($Q_n(p)$) in the vicinity of the critical point
is the product of the classical singular behavior multiplied
by an anomaly factor $a(n)$, which becomes extremely large
as $n \to \infty$ (and $p_c^n \to p_c$):
\begin{equation}
Q_n \sim a(n) (p/p_c^n - 1)^{\omega_{cl}} \ ,
\end{equation}
where $p$ is the control parameter and $\omega_{cl}$
is the classical critical index.
The divergence of this anomaly factor scales as
\begin{equation}
a(n) \sim (p_c^n - p_c)^{\omega - \omega_{cl}} \ ,
\end{equation}
thereby permitting the estimation of the true critical
exponent $\omega$, given a set of GMF approximation solutions.
The method has been tested on a great number of solved and
unsolved systems \cite{Suz-min}.
The application to cellular automata was shown by \cite{Odo}.
We followed the technique described there in the case of the
present CA models. A new parametrization is used suggested
by \cite{Kol}
\begin{equation}
\delta_n = (p_c/p_c^n)^{1/2} - (p_c^n/p_c)^{1/2} \ ,
\label{del}
\end{equation}
that has an invariance property: $p \leftrightarrow p^{-1}$.
We took into account a correction term to the anomaly scaling:
\begin{equation}
a(n) = b \ \delta_n^{\beta - \beta_{cl}} +
       c \ \delta_n^{\beta - \beta_{cl} + 1} \ .
\label{corr}
\end{equation}
The GMF calculation gave enough data to do the CAM
extrapolations for the one dimensional $y=1$ case (Table \ref{CAM1}).
In a similar way to the MFRG case for $y=2,3$ models the first
order predictions of the low level GMF results excluded CAM calculations.
Table \ref{CAM2} shows that $\beta$ results are stable
in contrast to neglecting points from the nonlinear fitting procedure
by Eq.(\ref{corr}) and indicating DP class behavior.

\section{Conclusions}

A deatiled analysis for a family of totalistic, critical CA in one and two
dimensions has been carried out in order to check the DP conjecture and to
clarify the former contradictory simulation results of the BBC ($y=3$) model.
Different critical exponents were determined by simple
steady state, finite size scaling and time dependent simulations.
The simulation results are more extensive than the previous ones and
none of them seems to break the DP rule. Rather they all give estimates
for different critical exponents that are close to the DP universality
values. We wish to emphasize that careful analysis is necessary
for each kind of simulation to get unambiguous results.
We have shown that the systematic calculation of GMF provides a
reliable way to distinguish the order of phase transitions.
Mean-field renormalization group and coherent anomaly method extrapolations
to GMF provided estimates for critical exponents. These results
are again in accordance with the DP conjecture.

\acknowledgments
The authors thank G. Szab\'o and N. Boccara for helpful
discussions. The simulations were carried out on the Connection
Machine-5 (grant no.: PHY930024N) and on the Fujitsu AP1000
supercomputers.
This research was partially supported by the Hungarian National
Research Fund (OTKA) under grant numbers T-4012 and F-7240.

\begin{figure}
\caption{Convergence of the $\beta$ exponent estimate in the scaling
region. Data are from steady state simulations.
The number of points corresponds to $p_{cut}$ such that the distance between
subsequent abscissa points is $\Delta p \sim 0.001$. (a) One dimension $y=1$.
(b) One dimension $y=2$. (c) One dimensions $y=3$. (d) Two dimensions $y=1$.}
\label{fig1}
\end{figure}

\begin{figure}
\caption{Concentration versus $p$ in two dimensions for (a) $y=2$.;
(b) $y=3$ models. Dashed lines correspond to GMF results of
$n=1,4$ point approximation. Data with error bars come from
steady state simulation}
\label{fig2}
\end{figure}

\begin{figure}
\caption{Finite size scaling of $c(L)$ and $\chi(L)$ for
the 1 dimensional BBC ($y=3$) model. For $y=1,2$ similar plots
were obtained. }
\label{fig3}
\end{figure}

\begin{figure}
\caption{Time dependent simulation results in one dimension for the $y=1$
model at $p$ values (from bottom to top) : $0.22086,\ 0.2209,\ 0.22092,\
0.22095,\ 0.22098$.
(a) The local slope of the logarithm of the $c(t)$ versus $\ln(1/t)$ is
plotted.
The intercept of the scaling curve with the ordinate gives exponent $\eta$.
(b) The local slope of the logarithm of the $s(t)$ versus $\ln(1/t)$ is
plotted.
The intercept of the scaling curve with the ordinate gives exponent $\delta$.}
\label{fig4}
\end{figure}

\begin{figure}
\caption{Time dependent simulation results in one dimension for the $y=2$
model at $p$ values (from bottom to top): $0.4354,\ 0.43545,\ 0.43547,\
0.4355,\ 0.4356$.
(a) The local slope of the logarithm of the $c(t)$ versus $\ln(1/t)$ is
plotted.
The intercept of the scaling curve with the ordinate gives exponent $\eta$.
(b) The local slope of the logarithm of the $s(t)$ versus $\ln(1/t)$ is
plotted.
The intercept of the scaling curve with the ordinate gives exponent $\delta$.}
\label{fig5}
\end{figure}

\begin{figure}
\caption{Time dependent simulation results in one dimension for the $y=3$
model at $p$ values (from bottom to top) : $0.7216,\ 0.7218,\ 0.7220$.
(a) The local slope of the logarithm of the $c(t)$ versus $\ln(1/t)$ is
plotted.
The intercept of the scaling curve with the ordinate gives exponent $\eta$.
(b) The local slope of the logarithm of the $s(t)$ versus $\ln(1/t)$ is
plotted.
The intercept of the scaling curve with the ordinate gives exponent $\delta$.}
\label{fig6}
\end{figure}

\mediumtext
\begin{table}
\caption{DP class critical exponents}
\begin{tabular}{cccc}
$\beta$    & $\eta$      & $\delta$    & $\nu_{\perp}$ \\
\tableline
$0.2767(4)$& $0.3137(1)$ & $0.1596(4)$ & $1.0972(6)$
\end{tabular}
\label{DP}
\end{table}

\mediumtext
\begin{table}
\caption{Simulation results in one dimension.}
\begin{tabular}{cccc}
       & $y=1$        & $y=2$        & $y=3$     \\
\tableline
TDMC  \\
$p_c$   & $0.22095(3)$ & $0.43543(5)$ & $0.7217(2)$ \\
$\eta$  & $0.317(7)$   & $0.317(7)$   & $0.307(9)$ \\
$\delta$& $0.149(5)$   & $0.159(7)$   & $0.161(8)$ \\
\tableline
STDMC \\
$p_c$  & $0.22103(4)$ & $0.435437(9)$  & $0.7198(1)$ \\
$\beta$& $0.267(2) $  & $0.258(1)$     & $0.298(7)$ \\
\end{tabular}
\label{pcbeta}
\end{table}

\mediumtext
\begin{table}
\caption{Small size effects in FSS as a funtion of cutoff for $y=1$ in one
dimension}
\begin{tabular}{cccccc}
cutoff & $0$ & $16$ & $32$ & $64$ & DP \\
\tableline
$\beta/\nu_{\perp}$&$0.260(4)$&$0.257(1)$&$0.255(1)$&$0.251(1)$&$0.2522(6)$ \\
$\gamma/\nu_{\perp}$&$0.43(1)$&$0.443(6)$&$0.462(6)$&$0.48(1)$&$0.496(2)$ \\
\end{tabular}
\label{fss1_6}
\end{table}

\mediumtext
\begin{table}
\caption{Small size effects in FSS as a funtion of cutoff for $y=2$ in one
dimension}
\begin{tabular}{ccccccc}
cutoff & $0$ & $16$ & $32$ & $64$ & $128$ & DP \\
\tableline
$\beta/\nu_{\perp}$&$0.252(1)$&$0.254(1)$&$0.254(2)$&$0.253(3)$&$0.252(4)$
&$0.2522(6)$ \\
$\gamma/\nu_{\perp}$&$0.548(7)$&$0.537(9)$&$0.52(1)$&$0.51(1)$&$0.50(2)$
&$0.496(2)$ \\
\end{tabular}
\label{fss2_6}
\end{table}

\mediumtext
\begin{table}
\caption{Small size effects in FSS as the funtion of cutoff for $y=3$ in one
dimension}
\begin{tabular}{cccccc}
cutoff & $0$ & $16$ & $32$ & $64$ & DP \\
\tableline
$\beta/\nu_{\perp}$&$0.258(1)$&$0.258(1)$&$0.255(1)$&$0.253(3)$&$0.2522(6)$ \\
$\gamma/\nu_{\perp}$&$0.496(4)$&$0.497(4)$&$0.493(7)$&$0.49(1)$&$0.496(2)$ \\
\end{tabular}
\label{fss3_6}
\end{table}

\narrowtext
\begin{table}
\caption{MFRG results for $y=1$ in one dimension}
\begin{tabular}{lrl}
$(n_i + n_{i+1})^{-1}$ & $p_c(i,i+1)$ & $\nu_{\perp}(i,i+1)$ \\
\tableline\tableline
$1/3$                  & $0.123106$   & $5.802$ \\
$1/5$                  & $0.135359$   & $3.287$ \\
$1/7$                  & $0.147078$   & $2.222$ \\
$1/9$                  & $0.157571$   & $1.627$ \\
\tableline
expected values        & simulation   & DP class \\
                       & $0.22095(3)$ & $1.0972(6)$ \\
\end{tabular}
\label{mfrg}
\end{table}

\mediumtext
\begin{table}
\caption{Convergence of the critical point estimates of
the one dimensional CA calculated by GMF approximation.
First order transitions are denoted by boldface numbers.
Gap sizes ($c(p_c)$ are shown for $y=2,3$.}
\begin{tabular}{cccccc}
$n$ & $y=1$ &\multicolumn{2}{c}{$y=2$} &\multicolumn{2}{c}{$y=3$} \\
    & $p_c$ & $p_c$ & $c(p_c)$ & $p_c$ & $c(p_c)$ \\
\tableline
$1$ & $0.111$ & ${\bf 0.354}$ & ${\bf 0.216}$ & ${\bf 0.534}$ & ${\bf 0.372}$\\
$2$ & $0.123$ & ${\bf 0.378}$ & ${\bf 0.228}$ & ${\bf 0.649}$ & ${\bf 0.385}$\\
$3$ & $0.135$ & ${\bf 0.401}$ & ${\bf 0.236}$ & $0.903$       &      $0.0$   \\
$4$ & $0.146$ & ${\bf 0.419}$ & ${\bf 0.215}$ & $0.800$       &      $0.0$   \\
$5$ & $0.157$ & ${\bf 0.427}$ & ${\bf 0.096}$ & $0.736$       &      $0.0$   \\
$6$ & $0.166$ & $0.426$       &        $0.0$  & $0.703$       &      $0.0$   \\
extrapolation & $0.196(6)$    & $0.467(7)$    & $0.0$         & $0.72\pm 0.1$ &
$0.0$ \\
\tableline
simulation& $0.22095(3)$ & $0.43543(5)$ & $0.0$ & $0.7217(2)$ & $0.0$ \\
\end{tabular}
\label{GMF1}
\end{table}

\mediumtext
\begin{table}
\caption{Convergence of the critical point estimates of
the two dimensional CA calculated by GMF approximation.
First order transitions are denoted by boldface numbers.
Gap sizes ($c(p_c)$ are shown for $y=2,3$.}
\begin{tabular}{cccccc}
$n$ & $y=1$ &\multicolumn{2}{c}{$y=2$} &\multicolumn{2}{c}{$y=3$} \\
    & $p_c$ & $p_c$ & $c(p_c)$ & $p_c$ & $c(p_c)$ \\
\tableline
$1$ & $0.111$ & ${\bf 0.354}$ & ${\bf 0.216}$ & ${\bf 0.534}$ & ${\bf 0.372}$
\\
$2$ & $0.113$ & ${\bf 0.326}$ & ${\bf 0.240}$ & ${\bf 0.455}$ & ${\bf 0.400}$\\
$4$ & $0.131$ & ${\bf 0.388}$ & ${\bf 0.244}$ & ${\bf 0.647}$ & ${\bf 0.410}$
\\
\tableline
simulation& $0.163$ & ${\bf 0.404}$ & ${\bf 0.245}$ & ${\bf 0.661}$ & ${\bf
0.418}$\\
\end{tabular}
\label{GMF2}
\end{table}

\narrowtext
\begin{table}
\caption{The basis of CAM fitting for $y=1$ in one dimension}
\begin{tabular}{ccc}
$n$ & $\delta_n$ & $a(n)$ \\
\tableline
$1$ & $0.701078$ & $0.248755$ \\
$2$ & $0.593013$ & $0.331049$ \\
$3$ & $0.494564$ & $0.428537$ \\
$4$ & $0.409735$ & $0.546581$ \\
$5$ & $0.339569$ & $0.673658$ \\
$6$ & $0.284204$ & $0.810893$ \\
\end{tabular}
\label{CAM1}
\end{table}

\narrowtext
\begin{table}
\caption{Stability of $\beta$ exponent for $y=1$ in one dimension calculated
by CAM}
\begin{tabular}{cc}
data set & $\beta$    \\
\tableline
$\{1,2,3,4,5,6\}$& $0.273$\\
$\{ \ \ ,2,3,4,5,6\}$& $0.280$ \\
$\{1 \ \ \ ,3,4,5,6\}$& $0.273$ \\
$\{1,2 \ \ \ ,4,5,6\}$& $0.283$ \\
$\{1,2,3 \ \ \ ,5,6\}$& $0.260$ \\
$\{1,2,3,4 \ \ \ ,6\}$& $0.273$ \\
$\{1,2,3,4,5 \ \ \ \}$& $0.266$ \\
\end{tabular}
\label{CAM2}
\end{table}

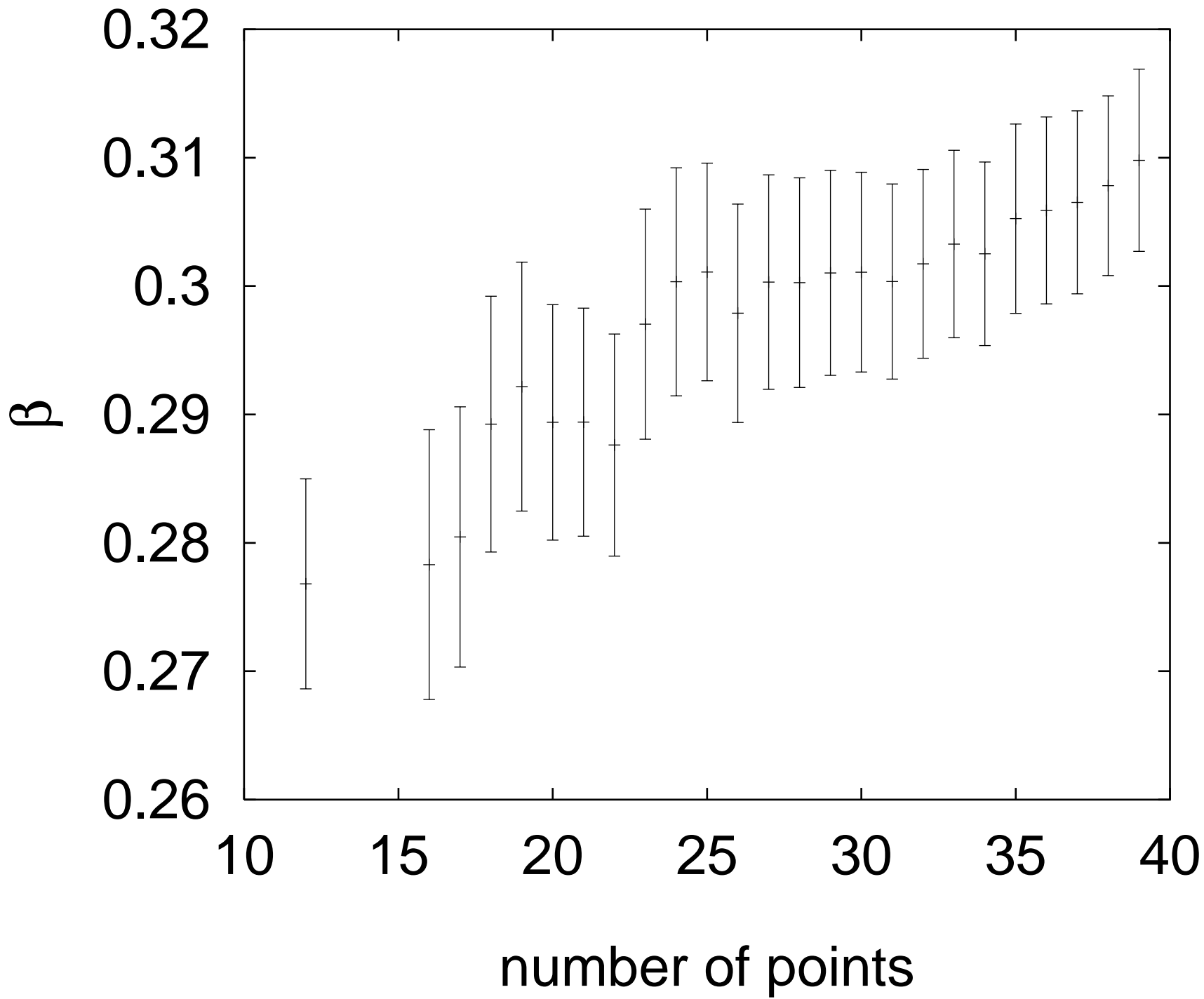
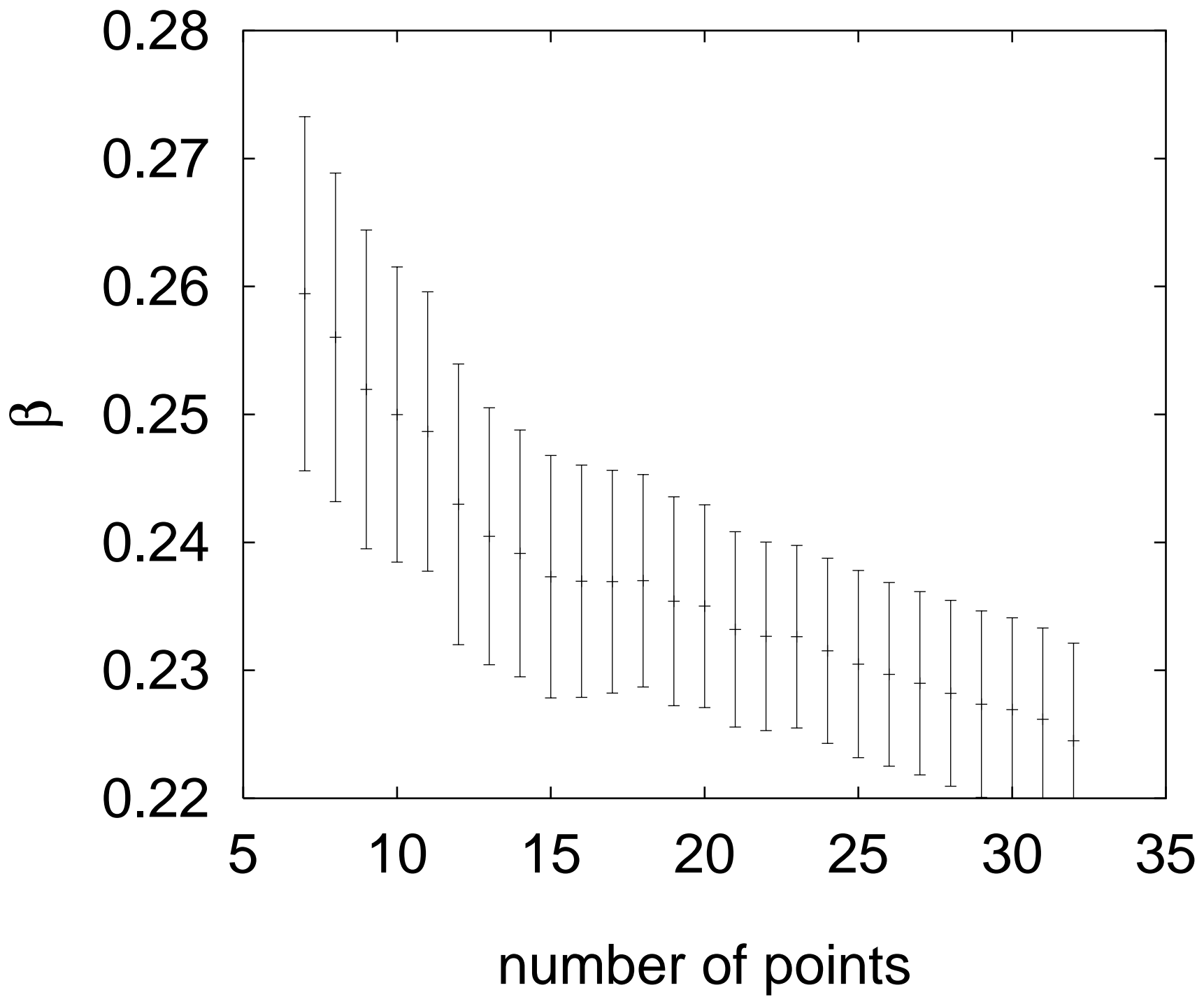
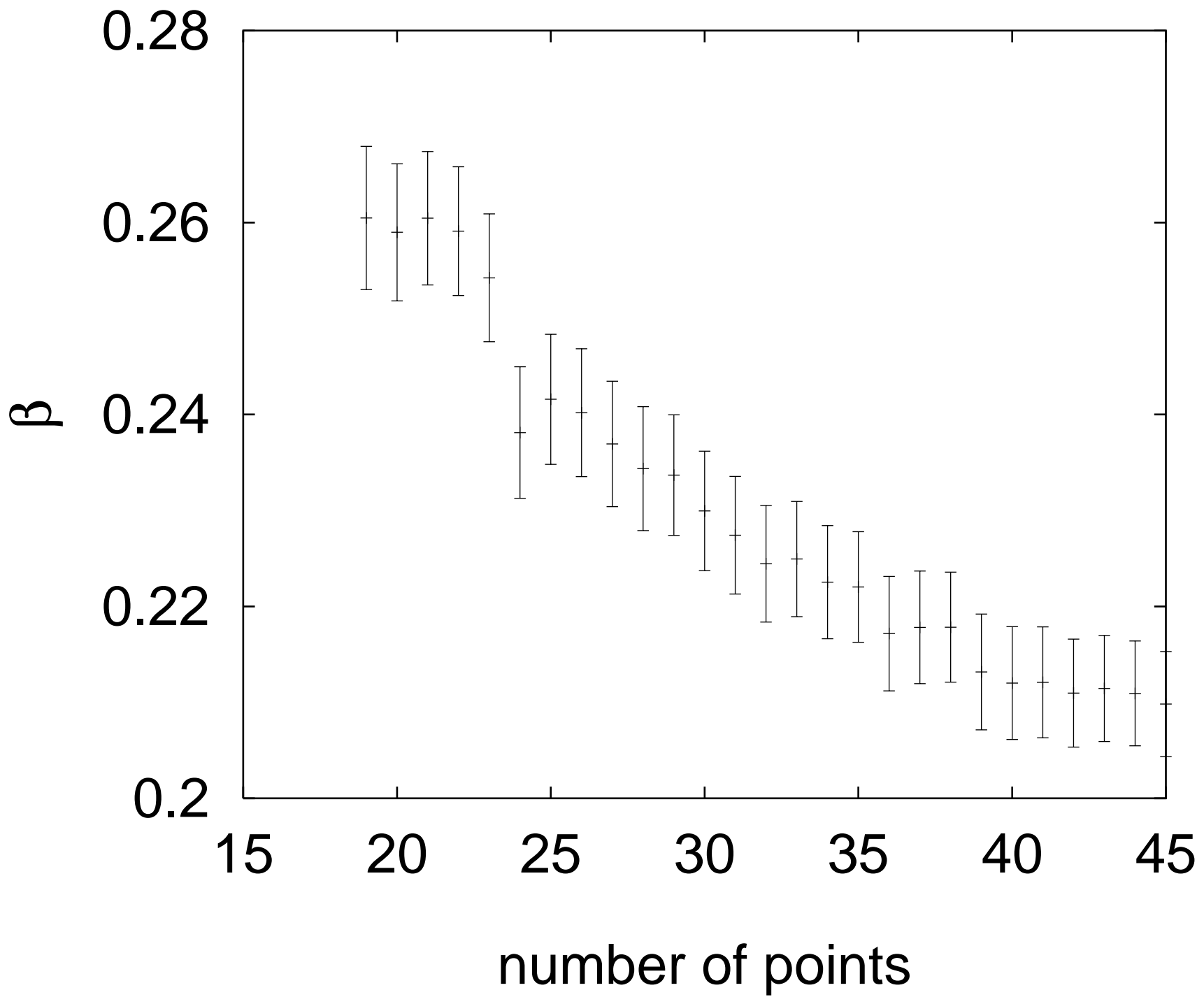
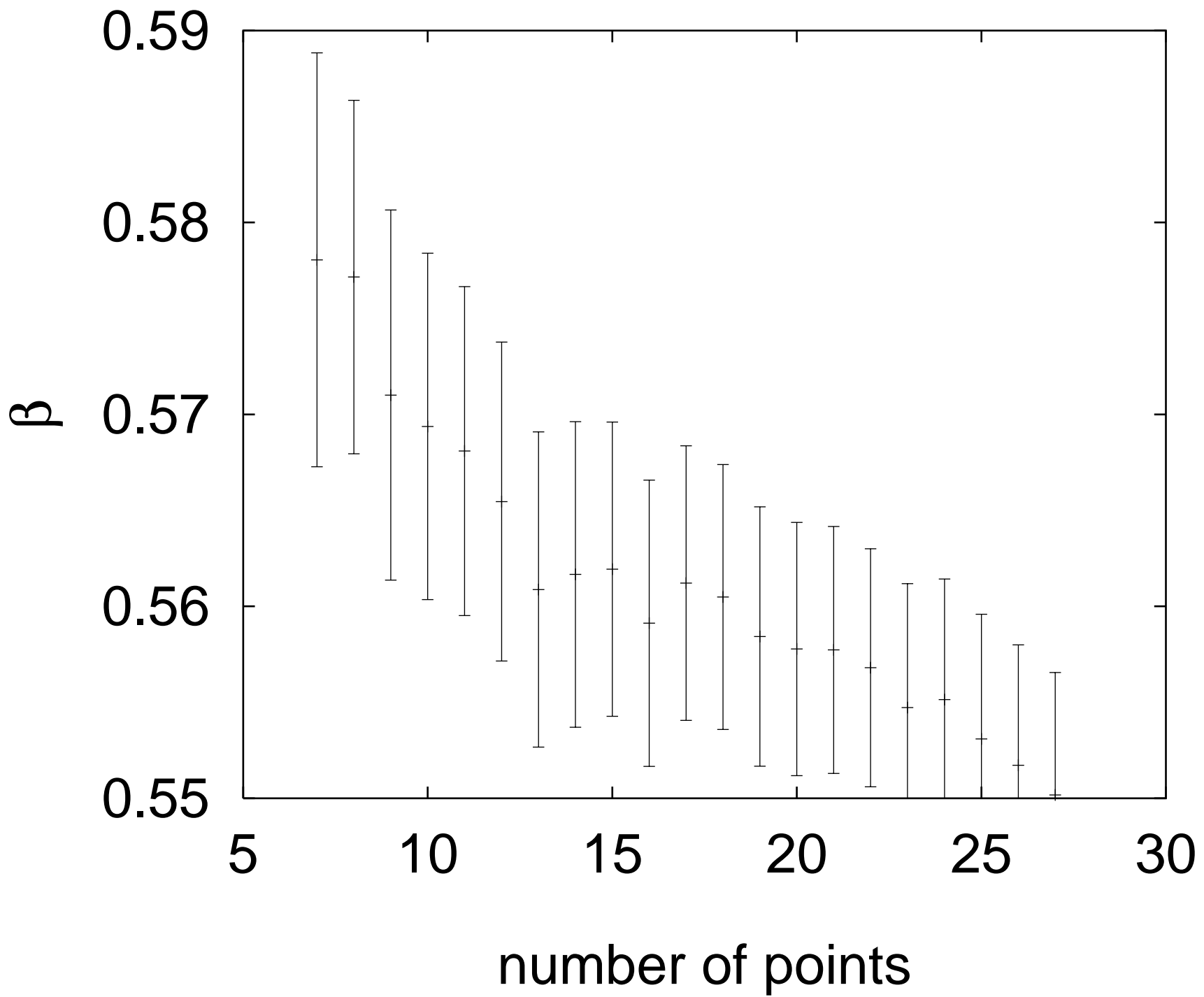
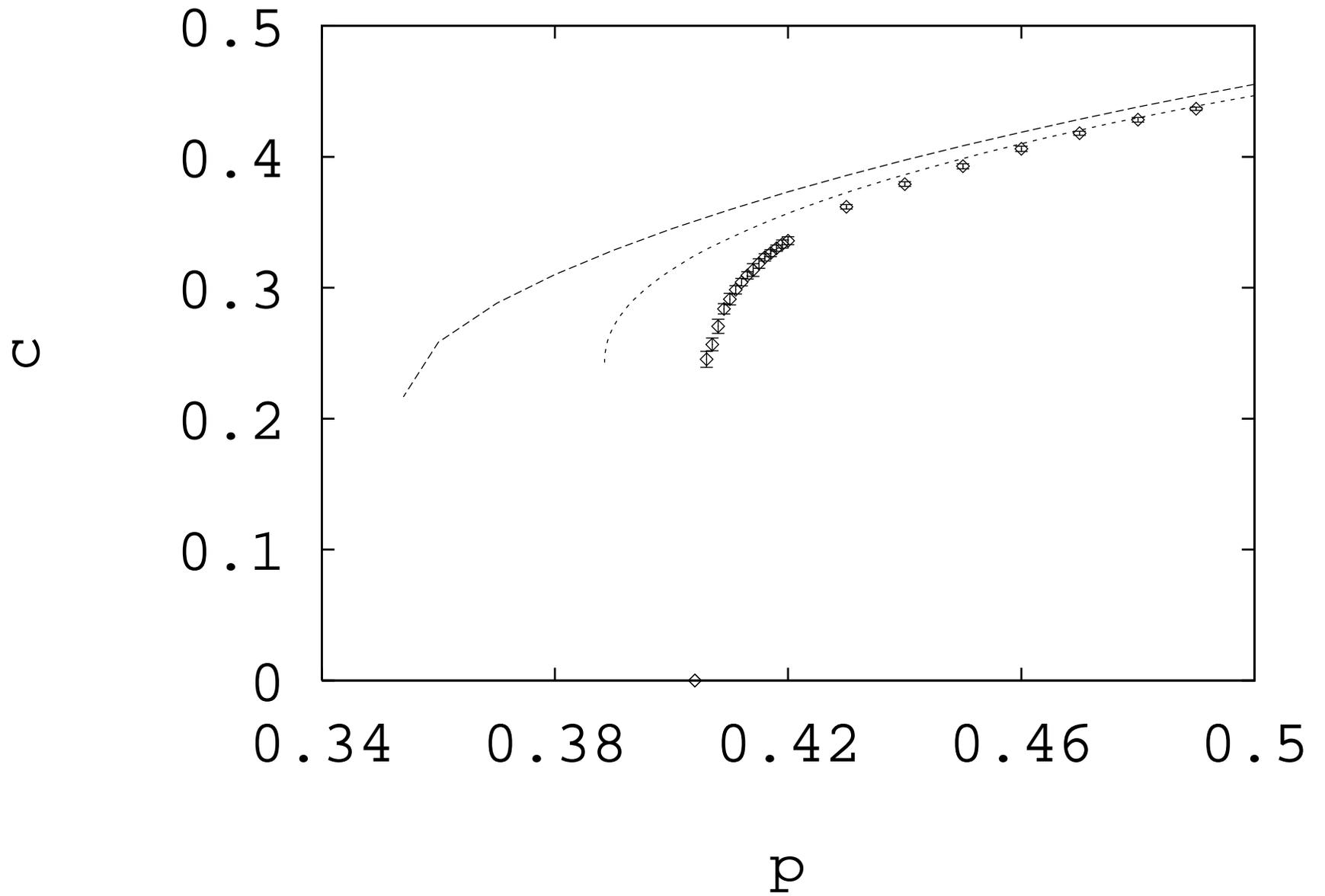
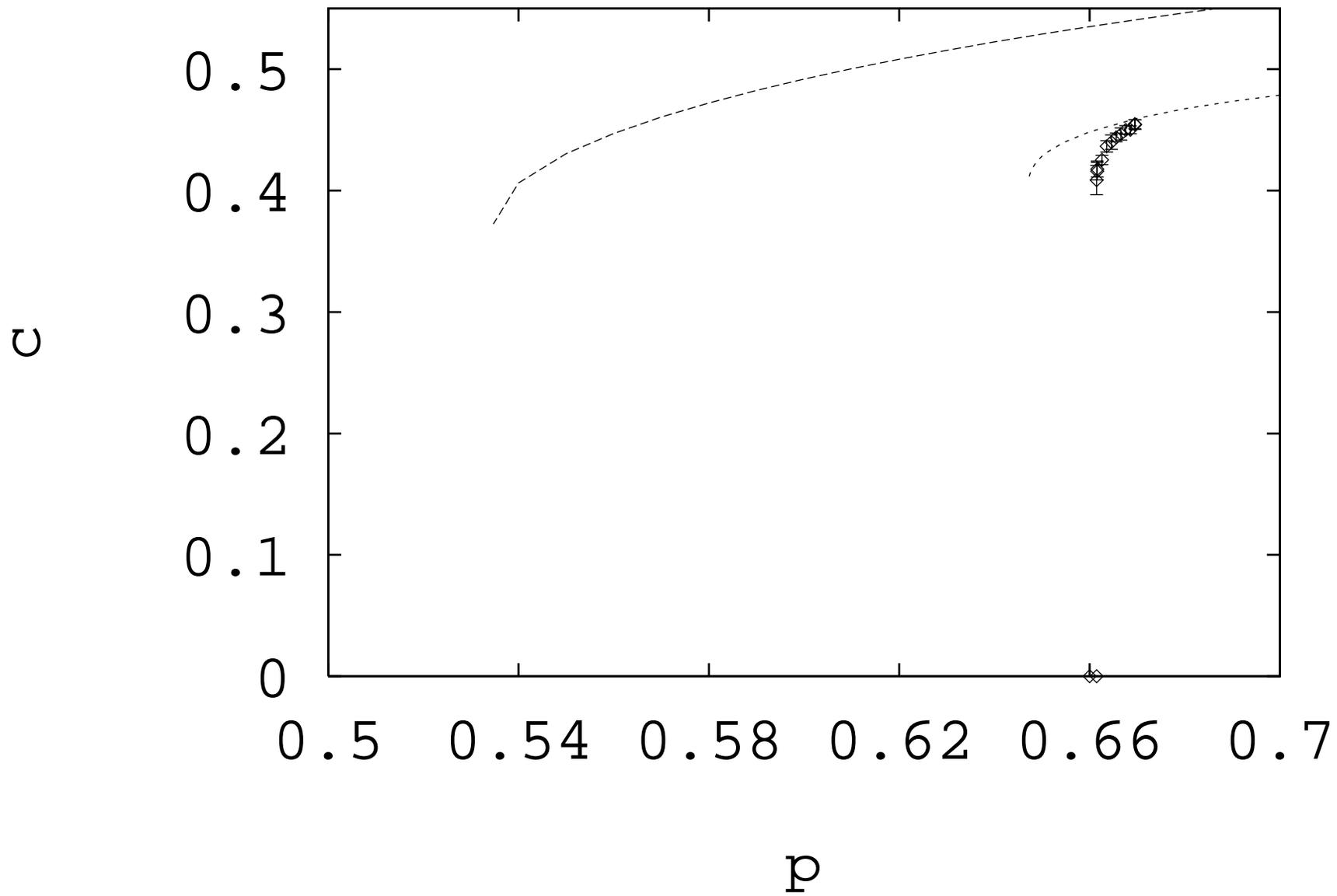
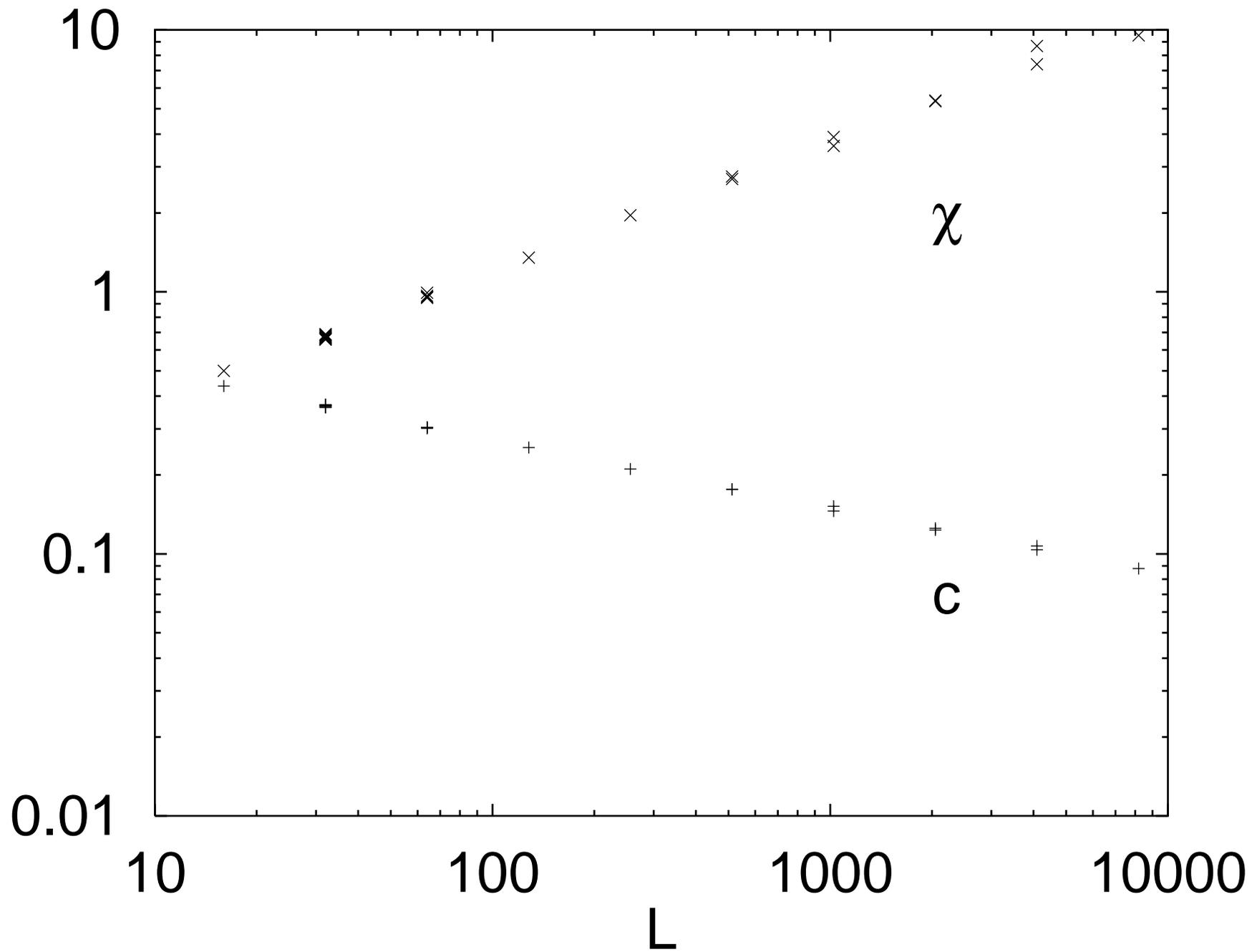
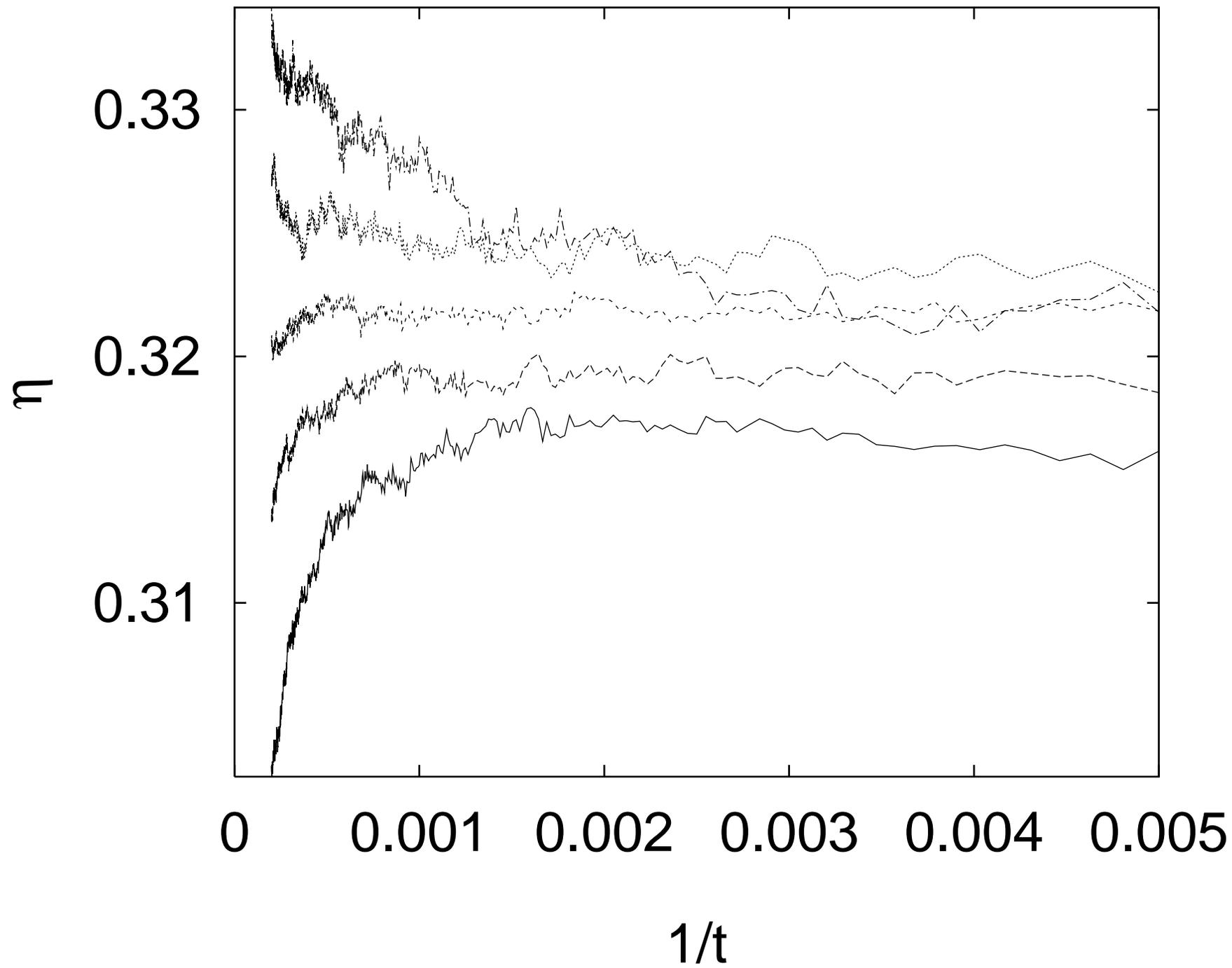
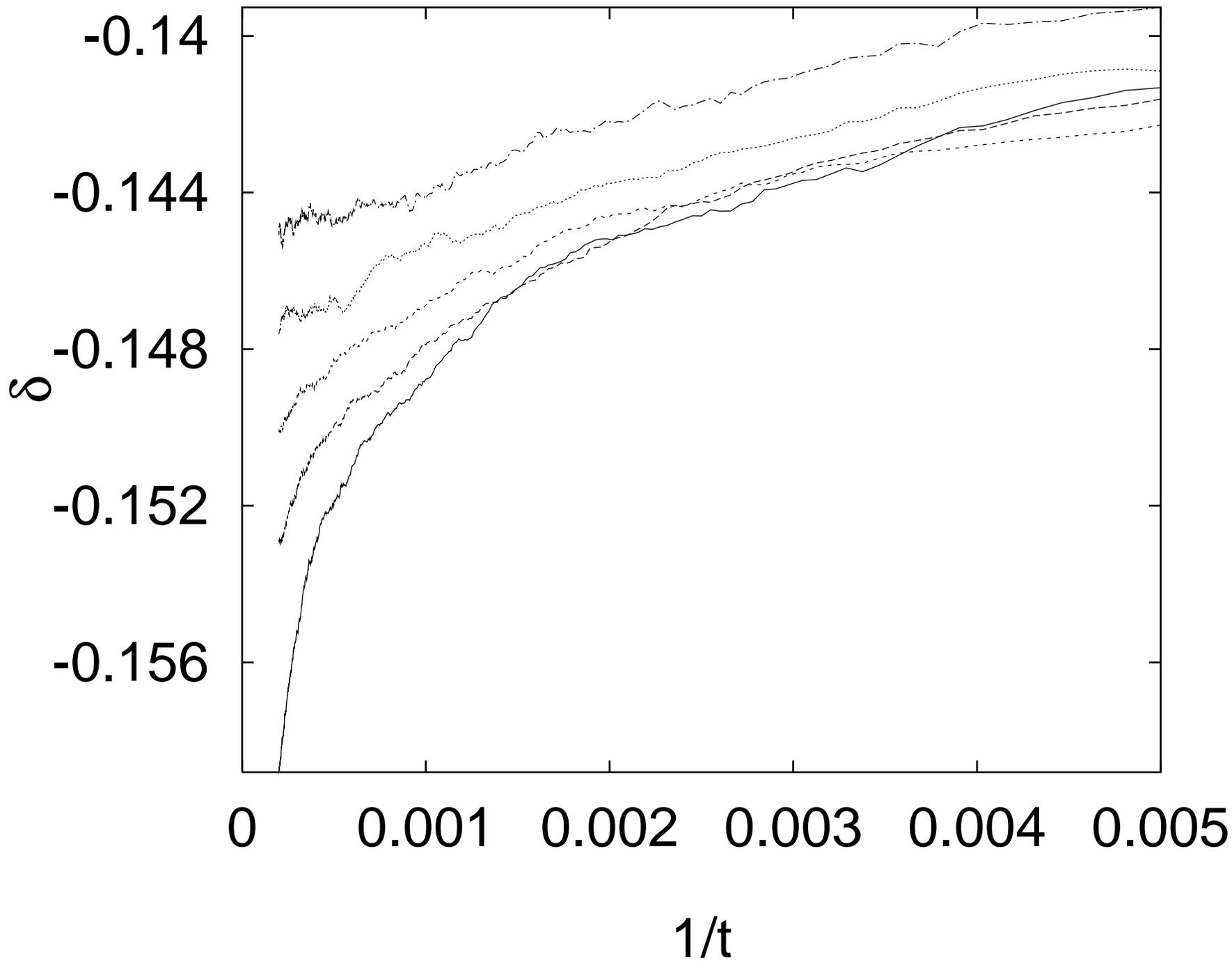
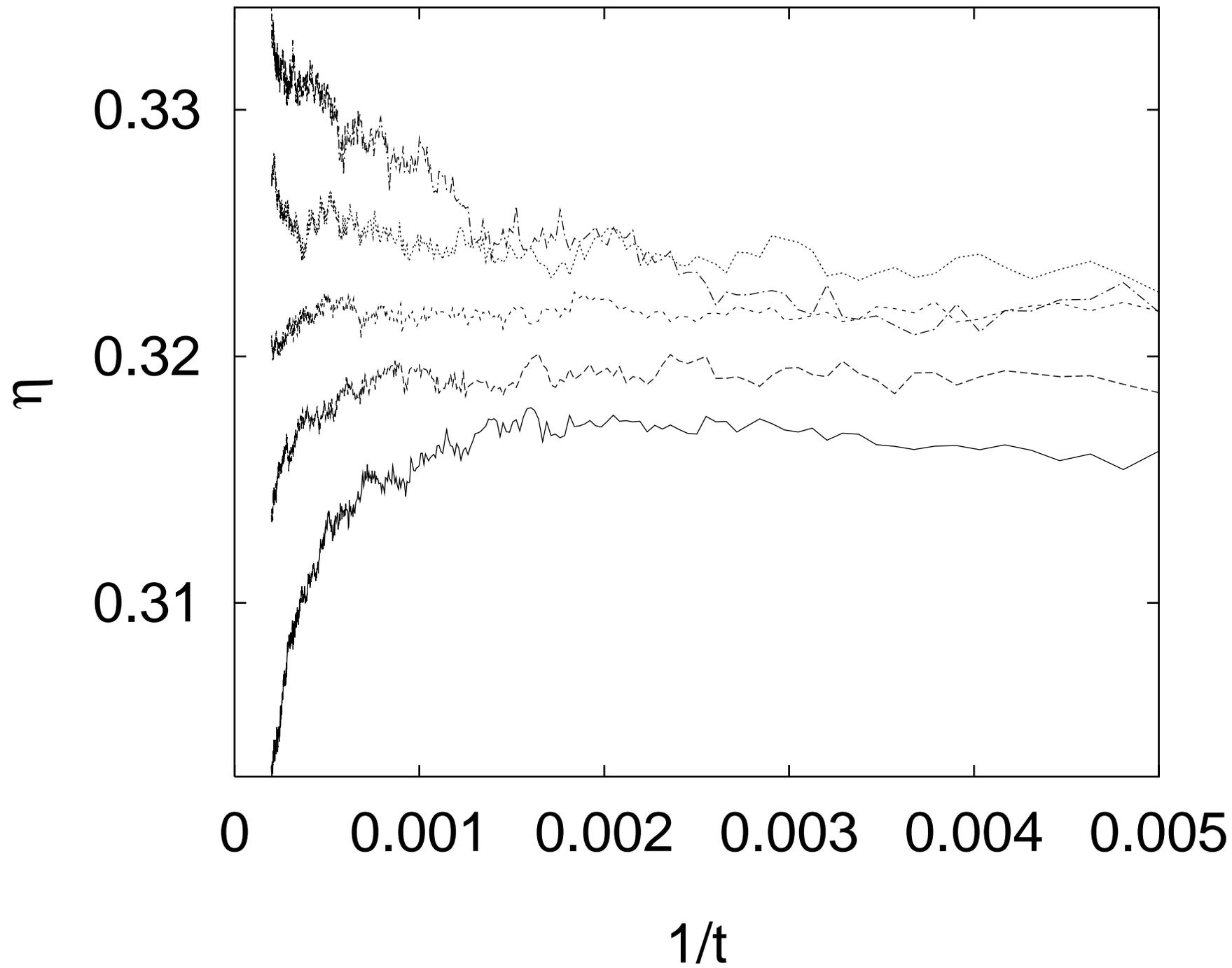
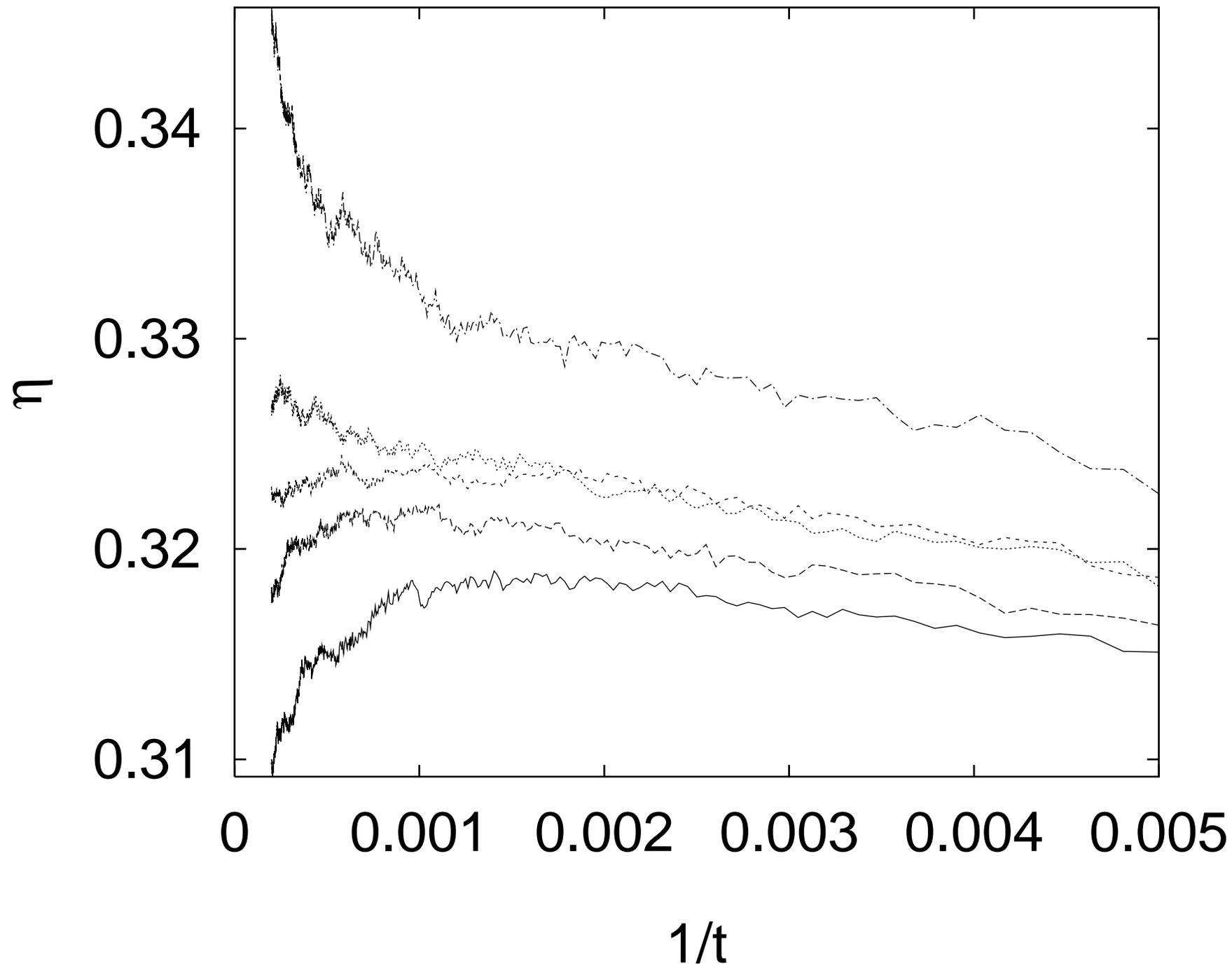
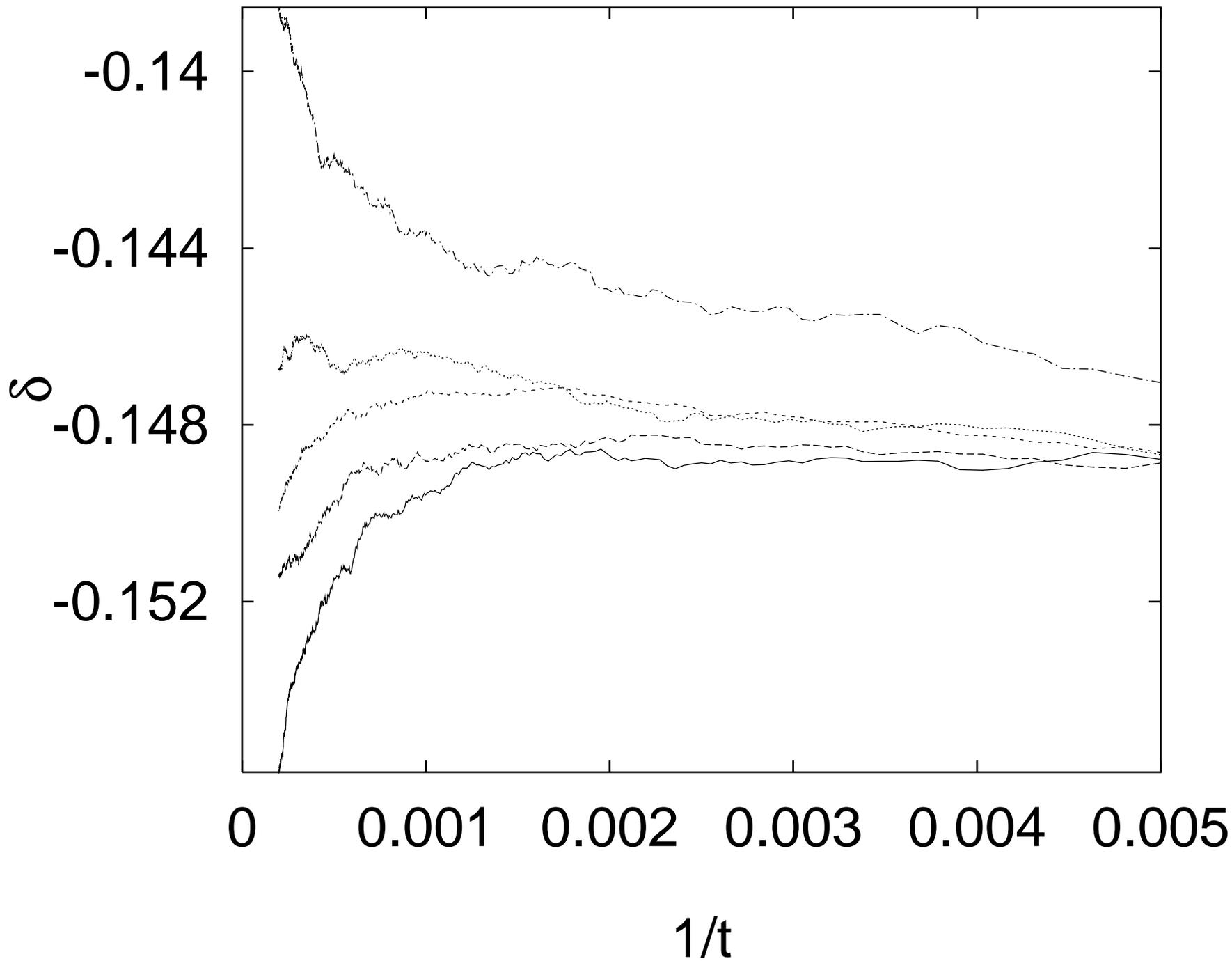
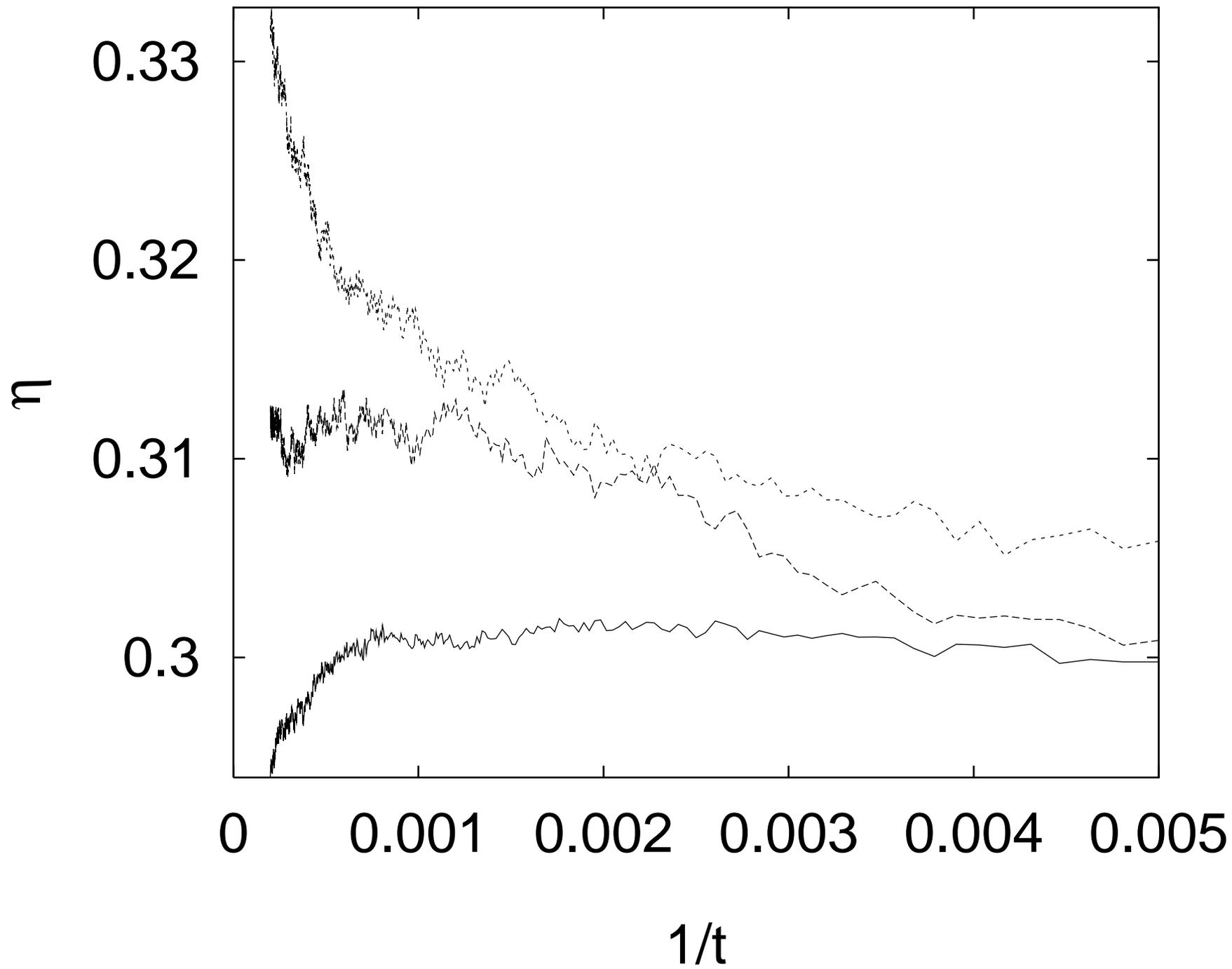
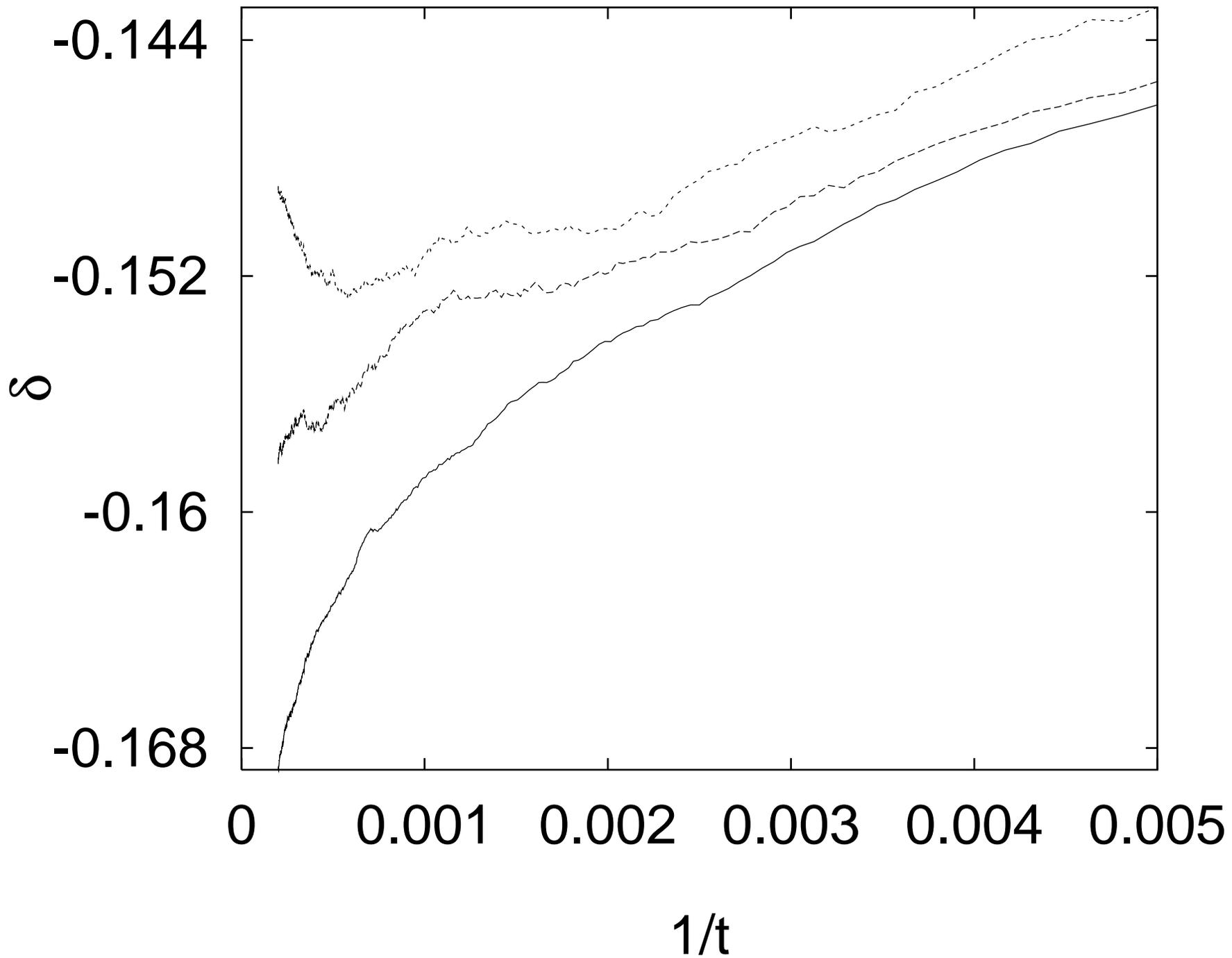
\end{document}